\documentclass[aps,prb,onecolumn,amsmath,10pt,groupedaddress,amssymb]{revtex4}

\usepackage{graphicx}   %
\usepackage{subfigure}  
\usepackage{hyperref}  

\newcommand{\ouN}{O\left(\frac{1}{N}\right)}
\newcommand{\dd}{\mathrm{d}}
\newcommand{\cN}{\mathcal{N}}

\DeclareMathOperator*{\tr}{Tr}

\begin{document}

\title{Finite size corrections to disordered systems on Erd\"{o}s-R\'enyi random graphs}

\author{U. Ferrari}
\affiliation{Laboratoire de Physique Th\'eorique de l'ENS, 
CNRS \& UPMC, 24 rue
Lhomond, 75005 Paris, France}
\author{C. Lucibello}
\affiliation{Dipartimento di Fisica, Universit\`a ``La Sapienza'', P.le A. Moro 2, I-00185, Rome, Italy}
\author{F. Morone}
\affiliation{Dipartimento di Fisica, Universit\`a ``La Sapienza'', P.le A. Moro 2, I-00185, Rome, Italy}
\author{G. Parisi}
\affiliation{Dipartimento di Fisica, INFN -- Sezione di Roma1, CNR-IPCF UOS Roma Kerberos, Universit\`a ``La Sapienza'', P.le A. Moro 2, I-00185, Rome, Italy}
\author{F. Ricci-Tersenghi}
\affiliation{Dipartimento di Fisica, INFN -- Sezione di Roma1, CNR-IPCF UOS Roma Kerberos, Universit\`a ``La Sapienza'', P.le A. Moro 2, I-00185, Rome, Italy}
\author{T. Rizzo}
\affiliation{CNR-IPCF, UOS Roma Kerberos, Dip. Fisica, Univ. ``La Sapienza'', P.le A. Moro 2, I-00185, Rome, Italy}

\begin{abstract}
We study  the finite size corrections to the free energy density in disorder spin systems on sparse random graphs, using both replica theory and cavity method. We derive an analytical expressions for the $O(1/N)$ corrections in the replica symmetric phase as a linear combination of the free energies of open and closed chains. We perform a numerical check of the formulae on the Random Field Ising Model at zero temperature, by computing finite size corrections to the ground state energy density. 
\end{abstract}

\maketitle

\section{Introduction}
The critical behaviour of ferromagnets in presence of a random magnetic field is  not well understood in spite of the great efforts that have been done in the past.  Dimensional reduction (i.e. the critical exponents of this system in $D$ dimensions are the same of a pure ferromagnet in $d=D-2$ dimensions) is perturbatively correct, but it fails beyond perturbation theory.  However it is not clear at the present moment if dimensional reduction is a valid approximation in some range of dimensions and which is the form of the deviations from dimensional reduction.
Different scenarios have been presented in the literature and they will not be discussed here:  we aim to  construct a new approach  to the problem.

The  difficulties are related to the following facts:
\begin{itemize}
\item The phase transition is dominated by the zero temperature fixed point:  the critical exponent as function of the temperature are the same as those as function of the magnetic field at zero temperature \cite{Bray1985}. 
\item The supersymmetric scenario (dimensional reduction)  assumes the essential uniqueness of the solution of the local mean field equations $m_i=\tanh(\beta h^{\text{\tiny{eff}}}_i)$ (at zero temperature they become $m_i=\mbox{sign}(h^{\text{\tiny{eff}}}_i)$, where  $h^{\text{\tiny{eff}}}_i\equiv \sum_{k}J_{ik}m_i+h_i$.
\end{itemize} 
The crux with the supersymmetry argument is than already at temperature higher than the critical temperature and certainly at zero temperature, the mean field equations have multiple solutions\cite{Lancaster1995}.

These  observations imply it would be wise to use a field theoretical approach directly at zero temperature, perturbing around  a mean field model where multiple solutions of the mean field equations are present. Unfortunately this is not immediate. The perturbation theory is usually constructed as an expansion around the mean field theory and the preferred mean field theory is the one for the infinite range model.  

In the infinite range model in the infinite volume limit the solution of the mean field equations is essentially unique (apart from a time reversal symmetry) \footnote{The only unknown of the model is the average magnetization of the system ($m$) that satisfies the equation $m=\int dP(h) \mbox{sign}( m+h)$.} and  we cannot perform any expansion around a non-existing transition with multiple solutions. However we must not throw out the baby with the bathwater. This disappointing situation disappears on the  Erd\"os-R\'enyi (ER) and other sparse random graphs, where the coordination number is finite and a more complex mean field theory is valid, where an exponential number of solutions is present (we may have many different solutions for the same value of the global magnetization \cite{NoglassRFIM}). 

The locality of the model on ER graphs, where the properties of a spin depend on the local magnetization averaged over its finite
neighbourhood, makes this problem deeply different from the infinite range model where only the global magnetization is relevant. Therefore we believe that the study of finite dimensional models performing an  expansion around the ER model is a mandatory investigation that may reserve us some surprises.

Our long term goal is to construct a new perturbation expansion around the ER graph results along the lines discussed in some previous works\cite{Parisi2012}\cite{Parisi2006}\cite{SackstederIV2013}. The construction of such a loop expansion for finite dimensional models is rather complex task. In this paper we present a first step in this direction, i.e. the study of the $1/N$ correction around the mean field solution for the ER graph. The tools that we use in this computations are the same of those that we should use in finite dimensions. Independently from this long term goal, the study of finite $N$ corrections is an interesting well studied problem, also because these corrections usually tell us something on the nature of the phase and the appearance of divergence in these corrections is often a signal of incorrectness of the mean field construction.

In the domain of physical spin systems, diluted models represent a class of mean-field like systems sharing an essential feature of the finite-dimensional ones, that is the finite coordination number. By consequence diluted models should mimic the physics of real systems better than the fully-connected ones (we have already remarked that this is what happens for zero temperature ferromagnets in random magnetic fields). Moreover when dealing with finite systems, the peculiar structure of diluted networks should give a first insight on how the topology can modify thermodynamic quantities. Indeed diluted models are defined on random graphs which are locally tree-like and have typical loops of size $O(\log N)$. However for finite (and small) sizes these loops become short and much more similar to the short loops which are abundant in any finite-dimensional network (think e.g.\ to lattice models). In this sense we can interpret the $1/N$ corrections in diluted models as a way to expand towards finite dimensional models.

Finite size corrections to the free energy have been investigated in fully connected systems \cite{PaSla93a,PaSla93,Campellone2009}, mean field optimization problems\cite{Parisi2002,Percus1996} and some simple disorder system\cite{Derrida1981}, sometimes as a byproduct of the Hessian diagonalization\citep{Mezard1987}. However, to our knowledge, only a solution in zero external field has been derived for sparse random graphs\cite{MontanariRizzo}\cite{Guerra2004} in the replica symmetric phase. 
In the following we will use the replica method in order to compute disorder-averaged corrections to the free energy. An obvious limitation of the method is that is cannot be applied on a given realization of the disorder to obtain corrections to the estimates provided by the Belief Propagation (BP) algorithm, which corresponds to the Bethe approximation. 
In order to tackle this problem a sequence of algorithms of increasing computational complexity have been proposed in \cite{MontanariRizzo} and it has been later tested that they indeed reduce systematically the error on the BP estimates \cite{Rizzo2007}. The sole limitation of these algorithms is that they 
do not give corrections to the free energy but only to local observables, notably the energy and the magnetization.

The paper is organized as follows. In Section II we define the model. In Section III we compute  finite size corrections of the free energy density in finitely connected models, using the replica formalism. In Section IV we make the same calculation using the cavity method. Since the cavity method is well defined only in the thermodynamic limit, it has to be reinvented in order to handle finite-size systems. We find that both procedures (replica \& cavity) give the same expression for the $1/N$ free energy density corrections and, in this respect, they are completely equivalent also beyond the thermodynamic limit. The cavity method allows a more precise physical interpretation of the finite-size corrections and of their connections with highly correlated topological structures (loops in the random graph). In Section V  we test our analytical predictions performing a numerical experiment on the zero-temperature Random Field Ising Model, by computing the $1/N$ corrections to the ground state energy. Numerical results are found to be in excellent agreement with the analytical prediction.

\section{The model}

We consider a model of $N$ interacting Ising spins $\{\sigma_{i}=\pm1\}_{i=1}^{N}$ defined by the following Hamiltonian:
\begin{equation}
\mathcal{H}=-\sum_{i<j}C_{ij}\,J_{ij}\sigma_{i}\sigma_{j}-\sum_{i}h_i\sigma_{i}\,\, ,
\label{hamilt}
\end{equation}
where we have decoupled the topology of the underlying graph, encoded in the symmetric adjacency matrix $\{C_{ij}\}$, from the exchange interactions $\{J_{ij}\}$. 
The numbers $C_{ij}$ specify the particular graph considered and take values
$C_{ij}=1\text{ or } 0$ whether the sites $i$ and $j$ are connected or not. Here
we consider Erd\"os-R\'enyi random graphs \cite{ERG}, which can be generated sampling
the adjacency matrix from the following distribution\cite{VianaBray}:
\begin{equation}
\mathcal{P}(\{C_{ij}\})=\prod_{i<j}\left[\frac{z}{N}\delta(C_{ij}-1)+\left(1-\frac{z}{N}\right)\delta(C_{ij})\right]\, .\label{eq:MatriceAdiacenza}
\end{equation}
The spins interact among each other via quenched random couplings $J_{ij}$,
which are assumed to be identically independently distributed (or fixed to a single value $J$). Moreover we allow the spins to interact with a local magnetic field (random or non-random).
The disorder averaged free energy density of the system, at the temperature $T=\beta^{-1}$, is defined as
\begin{equation}
f(\beta,N)=-(\beta N)^{-1}\left[\log Z_{N}(\beta)\right]_{\mathrm{av}}=f_0(\beta)+\frac{1}{N}f_1(\beta)+o\left(\frac{1}{N}\right),
\end{equation}
where the average has to be performed over the topological disorder
and the quenched randomness. The main part of this work is devoted to the analytical computation of the $f_1(\beta)$ term, the finite size correction to the free energy. The calculation can be performed in two different ways, known as the replica method and the cavity method. The latter derivation is particularly useful in order to better understand the physical meaning of the results, which is less clear in the replica picture.

\section{Computing the free energy density  with replicas}
The replica calculation of the free energy density starts from the well known identity:
\begin{equation}
\left[\log Z_{N}(\beta)\right]_{\mathrm{av}}=\lim_{n\rightarrow0}\ \frac{\partial}{\partial n}\log\left[(Z_{N}(\beta))^{n}\right]_{\mathrm{av}}\,\, .
\end{equation}
The moments of the partition function $\left[(Z_{N}(\beta))^{n}\right]_{\mathrm{av}}$ are then evaluated for integer values of number of replicas $n$. At the end of the calculation, the analytical continuation to real values of $n$ allows us to take the limit $n\rightarrow 0$. The replicated averaged partition function reads (from now on we drop the dependence of $Z_{N}$ on $\beta$): 
\begin{equation}
\left[(Z_{N})^{n}\right]_{\mathrm{av}}=\left[{\mathrm{Tr}}\left(\prod_{i<j}\exp\Big(\beta J_{ij}C_{ij}\sum_{a}^{n}\sigma_{i}^{a}\sigma_{j}^{a}\Big)\prod_{i}\exp\Big(\beta h_i\sum_{a}^{n}\sigma_{i}^{a}\Big)\right)\right]_{\mathrm{av}}\,\,. \label{eq:Zreplicata} 
\end{equation}
Performing the average over the topological disorder using the distribution \eqref{eq:MatriceAdiacenza}, and setting 
\begin{equation}
\begin{aligned}
&V(\sigma,\tau)\equiv N\log \left[1+\frac{z}{N}\left(\overline{\exp\Big(\beta J\sum_{a}\sigma^{a}\tau^{a}\Big)}^J -1\right)\right]\, ,\\
&B(\sigma)\equiv \log\left[\overline{\exp\Big(\beta h\sum_{a}\sigma^{a}\Big)}^{h}\right]-\frac{1}{2N}V(\sigma,\sigma)\, ,
\end{aligned}
\label{VU}
\end{equation}
eq. \eqref{eq:Zreplicata} takes the following form:
\begin{equation}
\left[(Z_{N})^{n}\right]_{\mathrm{av}}=\mathrm{Tr}\left[\exp\left\{\frac{1}{2N}\sum_{i,j}V(\sigma_i,\sigma_j)+\sum_iB(\sigma_i)\right\}\right]\, .\label{eq:Zreplicatamediatasulgrafo}
\end{equation}
We can achieve the site factorization of eq. \eqref{eq:Zreplicatamediatasulgrafo} by means of the order parameter
\begin{equation}
 \rho(\sigma)=N^{-1}\sum_i\prod_a\delta(\sigma^{a}-\sigma_{i}^{a})\,\, ,
 \label{constr}
\end{equation}
Enforcing eq. \eqref{constr} with a delta functional, integrating out the corresponding auxiliary field which appears in Gaussian form and taking the trace over the decoupled sites we arrive to an  expression suitable to saddle-point evaluation:
\footnote{The functional measure is $[D\rho]=\prod_\sigma\sqrt{\frac{N}{2\pi}}\dd\rho(\sigma)$}:
\begin{equation}
\left[(Z_{N})^{n}\right]_{\mathrm{av}}=\sqrt{\det(V)}\int[D\rho]e^{-NS[\rho]}\, \label{eq:funcInt}\,.
\end{equation}
The replicated action $S[\rho]$ is given by
\begin{equation}
S[\rho]=\frac{1}{2}\int \dd\sigma \dd\tau\ \rho(\sigma)\ V(\sigma,\tau)\rho(\tau)-\log\int \dd\sigma \exp\left[\int \dd\tau\ V(\sigma,\tau)\rho(\tau)+B(\sigma)\right]\,\, ,\label{eq:action}
\end{equation}
where the symbol ``$\int d\sigma$'' is a proxy for the more cumbersome  notation $\int \dd\sigma\equiv\prod_{a=1}^{n}\sum_{\sigma^a=\pm1}$\ .
Let us now extract the leading order contribution in the replicated action $S[\rho]$. We define the matrix $U(\sigma,\tau)$ and the vector $H(\sigma)$ from the first order expansion in $N$ of eq. \eqref{VU} to be
\begin{equation}
\begin{aligned}
U(\sigma,\tau)&\equiv\overline{\exp\Big(\beta J\sum_{a}\sigma^{a}\tau^{a}\Big)}^J \ ,\\
H(\sigma)&\equiv  \log\left[\overline{\exp\Big(\beta h\sum_{a}\sigma^{a}\Big)}^{h}\right]\ ,
\end{aligned}
\end{equation}
and write the thermodynamically relevant part of the action \eqref{eq:action} as $S[\rho]=S_0[\rho]+o(1)$, where
\begin{equation}
S_0[\rho]=\frac{z}{2}\int \dd\sigma \dd\tau\ \rho(\sigma)\ \big(U(\sigma,\tau)-1\big)\rho(\tau)-\log\int d\sigma \exp\left[z \int \dd\tau\ \big(U(\sigma,\tau)-1\big)\rho(\tau)+H(\sigma)\right]\,\, .
\label{s0}
\end{equation}
The leading order free energy $f_0$ comes from the saddle point of eq. \eqref{s0}, followed by the limit $n\to 0$, as we will see in the next section. A first $O\left(\frac{1}{N}\right)$ correction to the free energy comes from the $\ouN$ term in eq. \eqref{eq:action} evaluated at the saddle point. 

\subsection{Leading free energy}
We now evaluate the functional integral \eqref{eq:funcInt} by the steepest descent method:
\begin{equation}
\lim_{N\rightarrow +\infty}-\frac{1}{N}\log\left[(Z_{N})^{n}\right]_{\mathrm{av}}= S_0[\rho_*]\, ,
\end{equation} 
where $\rho_*(\sigma)$ is the solution of the the saddle-point equation: 
\begin{equation}
\frac{\delta S_0[\rho]}{\delta\rho(\sigma)}=0\quad\longrightarrow\quad\rho_*(\sigma)=
\frac{\exp\left[z\int \dd\ \sigma^{\prime}U(\sigma,\sigma^{\prime})\rho_*(\sigma^{\prime})+H(\sigma)\right]}{\int \dd\sigma\exp\left[z\int \dd\sigma^{\prime}U(\sigma,\sigma^{\prime})\rho_*(\sigma^{\prime})+H(\sigma)\right]}\, . \label{SP}
\end{equation}
In order to take the small $n$ limit we have to use an appropriate parametrization for the order parameter $\rho_*(\sigma)$. If we assume a Replica Symmetric (RS) ansatz, a convenient parametrization for $\rho_*(\sigma)$ is given by
\begin{equation}
\rho_*(\sigma)=\int dhP(h)\left[\frac{\exp\left(\beta h\sum_a\sigma^a\right)}{(2\mathrm{ch}(\beta h))^n}\right]\, .
\label{eq:RSparam}
\end{equation}
Inserting this parametrization in eq. \eqref{SP} and taking the limit $n\rightarrow0$ we obtain the usual self-consistent Cavity equations for the distribution $P(h)$ and $Q(u)$ of cavity fields and bias respectively:
\begin{equation}\begin{aligned}
&P(h)=\sum_{k=0}^{\infty}\frac{z^k}{k!}e^{-z}\ \overline{\int\left[\prod_{i=1}^{k}dQ(u_i)\right]\delta\left(h-h_R-\sum_{i=1}^{k}u_i\right)}^{h_R}\, \label{P(h)},\\
&Q(u)=\overline{\int dP(h)\ \delta\left[u-\frac{1}{\beta}\mathrm{th}^{-1}\left[\mathrm{th}(\beta J)\mathrm{th}(\beta h)\right]\right]}^{J}\, .
\end{aligned}
\end{equation}
The RS free energy density can then be estimated as
\begin{equation}
f_0(\beta)=\beta^{-1}\lim_{n\rightarrow 0}\ \frac{\partial}{\partial n}S_0[\rho_*]\, 
\end{equation}
and can be explicitly written in term of the distributions $P(h)$ and $Q(u)$ \cite{Mezard2009}.

\subsection{Fluctuations around the RS saddle point}
The Gaussian integral obtained by expanding eq.\eqref{eq:action} around the saddle point generates the order $1/N$ corrections. We set 
\begin{equation}
\begin{aligned}
&\rho(\sigma)=\rho_*(\sigma)+\frac{\chi(\sigma)}{\sqrt{N}}\, , \\
&S^{(2)}(\sigma,\sigma^{\prime};\rho)=\frac{\delta^{2}S_0[\rho]}{\delta\rho(\sigma)\delta\rho(\sigma^{\prime})}\, .
\end{aligned}
\end{equation}
Expanding the action in powers of $1/N$ we find

\begin{equation}
S[\rho]=S_0[\rho_*]+\frac{1}{N}S_1[\rho_*]+\frac{1}{2N}\int d\sigma d\sigma^{\prime}\chi(\sigma)S^{(2)}(\sigma,\sigma^{\prime};\rho_*)\chi(\sigma^{\prime})+o(N^{-1})\, ,
\end{equation}
where $S_1[\rho_*]$ is given by the following expression:
\begin{align}
S_1[\rho_*] =\frac{z}{2}\int\dd\sigma\left[(U(\sigma,\sigma)-1\right]\rho_*(\sigma)+\frac{z^2}{4}\int\dd\sigma\dd\sigma' \rho_*(\sigma)\left[U(\sigma,\sigma')-1\right]^2\rho_*(\sigma')\ .\label{eq:S1}
\end{align}
The functional integral \eqref{eq:funcInt} at this order evaluates:
\begin{equation}
\begin{aligned}
-\frac{1}{N}\log\left[(Z_{N})^{n}\right]_{\mathrm{av}} 
=\ & S_0[\rho_*] + \frac{1}{N}S_1[\rho_*]+\frac{1}{2N}\log\det\left(1-T\right) +o(N^{-1})\\
=\ & S_0[\rho_*] + \frac{1}{N}S_1[\rho_*]-\frac{1}{2N}\sum_{L=1}^{\infty}\frac{\mathrm{Tr}[T^L]}{L}+o(N^{-1})\, ,
\end{aligned}
\end{equation}
where the matrix $T(\sigma,\sigma^{\prime})$ reads
\begin{equation}
T(\sigma,\sigma^{\prime})=z\left[U(\sigma,\sigma^{\prime})\rho_*(\sigma^{\prime})-\left(\int d\tau U(\sigma,\tau)\rho_*(\tau)\right)\rho_*(\sigma^{\prime})\right]\, .
\end{equation}
Using the RS parametrization \eqref{eq:RSparam}, it turns out that in the limit $n\downarrow 0$ the trace $\tr\left(T^L\right)$ can be arranged in a linear combination of free energies of closed and open chains. It all comes down to the fact that the term $U(\sigma,\sigma')\rho_*(\sigma')$, present in $T(\sigma,\sigma')$, can be linked to the replicated transfer matrix of an edge receiving a cavity field at one of its extremities. In Appendix \ref{app:combin} we prove the following formula:
\begin{equation}
\frac{\partial}{\partial n}\mathrm{Tr}(T^L)=-\beta z^L\left[\phi_L^c-L\left(\phi_L^a-\phi_{L-1}^a\right)\right]+O(n)\, ,\label{eq:tracciacatene}
\end{equation} 
where $\phi_L^{c/a}$ are \textbf{free energies} of closed and open spin chains in the graph of length $L\geq1$, with $\phi_0^a$ defined as $\phi_0^a\equiv-\beta^{-1}\mathbb{E}_h\log2\cosh(\beta h)$. Writing the RS free energy density as:
\begin{equation}
f_{\mathrm{RS}}=f^{(0)}_{\mathrm{RS}}+\frac{1}{N}f^{(1)}_{\mathrm{RS}}+o\left(N^{-1}\right)\, ,
\label{fRS}
\end{equation}
and observing that the term $S_1[\rho_*]/N$ [viz. eq. \eqref{eq:S1}] cancels out with part of the first two terms in the sum $\sum_{L=1}^{\infty}\text{Tr}\left(T^L\right)/(2NL)$,
the finite size correction of the RS free energy density  $f^{(1)}_{\mathrm{RS}}$ can be evaluated as: 
\begin{equation}
\boxed{
f^{(1)}_{\mathrm{RS}}=\left(z-\frac{z^2}{2}\right)\phi_0^a-\frac{z}{2}\phi_1^a-\frac{z^2}{2}(\phi_2^a-2\phi_1^a)+\frac{1}{2}\sum_{L=3}^{\infty}\frac{z^L}{L}\left[\phi_L^c-L(\phi_L^a-\phi_{L-1}^a)\right]}\,  .
\label{eq:F1}
\end{equation}
The sum entering the previous formula can be considered as a sum over independent loops weighted with the factor $\left[\phi_L^c-L(\phi_L^a-\phi_{L-1}^a)\right]$, by noticing that, in the thermodynamic limit, $z^L/(2L)$ is exactly the average number of loops of length $L$ in a Erd\"{o}s-R\'enyi random graph of mean connectivity $z$. The same formula holds true also on the Erd\"{o}s-R\'enyi ensemble $\mathbb{G}(N,M)$, where $M=z N / 2$ is the fixed number of edges, since the distribution of topological structures such as the number of finite loops remains the same at the $1/N$ order. \\
In the limit of vanishing external field, eq.\eqref{eq:F1}, evaluated in the paramagnetic phase, takes the following simpler form:
\begin{equation}
f^{(1)}_{\mathrm{RS}}=\frac{z}{2\beta}\mathbb{E}_J\log\mathrm{ch}(\beta J)- \frac{1}{2\beta}\sum_{L=3}^{\infty}\frac{z^L}{L}\mathbb{E}_{\{J_i\}}\log\left[1+\prod_{i=1}^{L}\mathrm{th}(\beta J_i)\right]\, ,\label{eq:fh0}
\end{equation}  
where the first term takes into account the fact that the average number of links is $z(N-1)/2$ and the second one is the contribution of all the loops of length $L\geq3$. The loops we are talking about are topologically defined as non-self-intersecting closed paths. Self-intersecting closed paths would give contributions proportional to $N^{-2}$, since the self-intersection is observed, on average, in a fraction ${N^{-2}}$ of the total number of vertices. While eq. \eqref{eq:F1} is an original contribution to the literature, its zero field counterpart eq. \eqref{eq:fh0} has been already presented \cite{MontanariRizzo}. Moreover the full distribution of  $f^{(1)}$ in the absence of external field and in the RS phase has been rigorously computed\cite{Guerra2004} an it is consistent with the mean value given by eq. \eqref{eq:fh0}.

\section{Computing the free energy density  with cavity method}

We now show how to compute the finite size corrections to the free energy density using the cavity method. The reason to be interested in such a kind of calculation is twofold. Firstly we have to corroborate the physical insight gained from replicas; secondly we want to establish the equivalence of the two methods beyond the leading order, showing how both procedures give the same result also at order $1/N$.

The cavity method is well defined only in thermodynamic limit. In order to study $1/N$ corrections to the free energy density of a model defined on a Erd\"os-R\'enyi random graph (ERRG), we need to define a new ensemble of random graphs of $\mathcal{N}$ vertices, such that in the limit $\mathcal{N}\rightarrow\infty$ any topological structure appears with the same density it has in the ERRG of $N$ vertices.
Here we are assuming that the free energy of a model of $N$ variables can be written as $F_N = N f(\{d_i\})$, where $f(\{d_i\})$ is the free energy density computed in the thermodynamic limit on a model having the same \emph{densities} $d_i$ of topological structures appearing in the finite $N$ model. The new ensemble we are going to define is required to compute such a free energy density.

The topological structures we are interested in are the only ones that give contributions up to order $\ouN$, i.e linear chains of length $L$ (i.e. with $L$ edges and $L+1$ vertices) and loops of length $L$. Let us start by computing their densities in a ERRG of $N$ sites, where each link is present with probability $z/N$. The density of linear chains of size $L$ (i.e. the number of linear chains per node) is 
\begin{equation}
d_{L}^{chain}=\frac{1}{N}\left(\frac{z}{N}\right)^{L}\frac{1}{2}N(N-1)\dots(N-L)\simeq\frac{z^{L}}{2}\left(1-\frac{L(L+1)}{2N}\right)\, ,
\label{eq:dens_chains}
\end{equation}
and the density of loops of length $L$ is 
\begin{equation}
d_{L}^{loop}=\frac{1}{N}\left(\frac{z}{N}\right)^{L}\frac{1}{2L}N(N-1)\dots(N-L+1)\simeq\frac{1}{N}\frac{z^{L}}{2L}\, .
\label{eq:dens_loops}
\end{equation}

In the new ensemble a random graph of $\mathcal{N}$ nodes can be viewed as the union of basic topological structures (BTS), that, for the present purposes, are chains and loops.
The graph can be build in the following way.
For each $L \ge 1$, consider all sequences of $L+1$ different indices $(i_0,i_1,\ldots,i_L)$ with the condition $i_0<i_L$, that avoids double counting of a chain; for each sequence of indices draw the edges $(i_0,i_1), (i_1,i_2), \ldots, (i_{L-1},i_L)$ with probability $a_L/\mathcal{N}^{L}$.
Then, for each $L \ge 3$, consider all sequences of $L$ different indices $(i_1,i_2,\ldots,i_L)$ with the conditions that $i_1$ is the smallest among the $L$ indices and $i_2<i_L$ (these two conditions ensure that each loop is counted only once); for each sequence of indices draw the edges $(i_1,i_2), (i_2,i_3), \ldots, (i_{L-1},i_L), (i_L,i_1)$ with probability $c_L/\mathcal{N}^{L-1}$.

A useful representation of this graph is in terms of a factor graph, where the variable nodes are the graph nodes and the factor nodes are the BTS. Thanks to the scaling of the probabilities used in the building of the graph, the corresponding factor graph is sparse, since the total number of BTS (i.e., of factor nodes) is given by
\[
\sum_{L=1}^\infty \frac{\cN(\cN-1)\ldots(\cN-L)}{2} \frac{a_L}{\cN^L} +
\sum_{L=3}^\infty \frac{\cN(\cN-1)\ldots(\cN-L+1)}{2L} \frac{c_L}{\cN^{L-1}} \simeq
\cN \left( \sum_{L=1}^\infty \frac{a_L}{2} + \sum_{L=3}^\infty \frac{c_L}{2L} \right)
\]
and the coefficients $a_L$ and $c_L$ are constants.

The sparsity of the factor graph ensures that the whole construction is consistent in the $\mathcal{N} \to \infty$ limit. Indeed the probability that any pair of graph nodes enters in more than one BTS is $O(1/\cN)$. Since in the new ensemble we are interested in computing the free energy density to leading order, we can safely assume that any two graph nodes interact through at most one BTS; and this BTS uniquely determines whether the edge between the two graph nodes is present or not.

The factor graph representation also allows us to write down the free energy density in a standard way by summing factor nodes and variable nodes contributions
\begin{equation}
f=\frac{1}{2}\sum_{k=1}^{\infty}a_k\phi_{k}^{a}+\frac{1}{2}\sum_{k=3}^{\infty}\frac{c_k}{k}\phi_{k}^{c}+\phi_{site}\, ,
\label{eq:cavityF}
\end{equation}
where $\phi_k^a$ and $\phi_k^c$ are respectively the free energies of chains and loops of length $k$ and
\[
\phi_{site}=\frac{T}{\cN}\sum_{i}(1-n_i)\sum_{\sigma_{i}}\mu_i(\sigma_{i})\log \mu_i(\sigma_{i})\;,
\]
with $\mu_i(\sigma_{i})$ being the single spin marginal and $n_i$ the number of BTS where the variable $i$ enters.

We should now determine the values of the coefficients $a_k$ and $c_k$ such that the densities of chains and loops in a typical graph of the new ensemble match those in eqs. (\ref{eq:dens_chains}) and (\ref{eq:dens_loops}) in the large $\cN$ limit.
When computing the actual density of a given topological structure (e.g. a chain or a loop) one should consider that such a topological structure can coincide with a BTS, or be part of a BST or involve more than one BST.

As a warm-up, let us compute the density of links (chains of length $L=1$) in the limit $\mathcal{N}\rightarrow\infty$:
\begin{equation}
d_{1}^{chain}=\lim_{\mathcal{N}\rightarrow\infty}\frac{1}{\mathcal{N}}\frac{{\mathcal{N}}^2}{2}\left[\sum_{k=1}^{\infty}k{\mathcal{N}}^{k-1}\frac{a_k}{{\mathcal{N}}^k}+\sum_{k=3}^{\infty}{\mathcal{N}}^{k-2}\frac{c_k}{{\mathcal{N}}^{k-1}}\right]=\frac{1}{2}\left[\sum_{k=1}^{\infty}ka_k+\sum_{k=3}^{\infty}c_k\right]\, ,
\end{equation}
where $k{\mathcal{N}}^{k-1}$ in the first sum and ${\mathcal{N}}^{k-2}$ in the second sum are respectively the number of chains and loops of length $k$ passing trough a given link, i.e., the number of possible BTS containing the two variables connected by a given link.

When computing the density of topological structures made of more than one link, we need to consider that such structures can overlap with more than one BTS. In order to be concrete let us consider the density of chains of length $L=2$:
\begin{equation}
d_{2}^{chain}=\lim_{\mathcal{N}\rightarrow\infty}\frac{1}{\mathcal{N}}\frac{{\mathcal{N}}^3}{2}\left[\left(\frac{2d_{1}^{chain}}{\mathcal{N}}\right)^2+\sum_{k=2}^{\infty}(k-1){\mathcal{N}}^{k-2}\frac{a_k}{{\mathcal{N}}^k}+\sum_{k=3}^{\infty}{\mathcal{N}}^{k-3}\frac{c_k}{{\mathcal{N}}^{k-1}}  \right]\, 
\end{equation}
where $2d_{1}^{chain}/\mathcal{N}\equiv p_1$ is the probability of having a link\footnote{In principle in $p_1$ there are already some contributions entering the sums, but this over-counting effect is irrelevant in the $\mathcal{N}\rightarrow\infty$ limit.}. The general expression for densities of linear chains of length $L\geq 3$ is the following
\begin{equation}
d_{L}^{chain}=\lim_{\mathcal{N}\rightarrow\infty}\frac{1}{\mathcal{N}}\frac{{\mathcal{N}}^{L+1}}{2}\left[S_L\left(\frac{2d_{1}^{chain}}{\mathcal{N}},\dots,\frac{2d_{L-1}^{chain}}{{\mathcal{N}}^{L-1}}\right)+\sum_{k=L}^{\infty}(k-L+1)\frac{a_k}{{\mathcal{N}}^L}+\sum_{k=L+1}^{\infty}\frac{c_k}{{\mathcal{N}}^{L}} \right]\, ,
\end{equation}
where the function $S_L$ gives the probability that the $L$ consecutive links comes from more than one BTS and can be written (see Appendix A) in terms of the probabilities of having $k(<L)$ consecutive links: $p_k \equiv 2d_{k}^{chain}/\mathcal{N}^k$. Since each term in function $S_L$ is of order ${\mathcal{N}}^{-L}$, in the limit ${\mathcal{N}}\rightarrow\infty$ we have
\begin{multline}
2d_{L}^{chain} = \cN^L S_L(p_1,\dots,p_{L-1})+\sum_{k=L}^{\infty}(k-L+1)a_k+\sum_{k=L+1}^{\infty}c_k =\\
S_L(2d_1^{chain},\dots,2d_{L-1}^{chain})+\sum_{k=L}^{\infty}(k-L+1)a_k+\sum_{k=L+1}^{\infty}c_k = z^{L}\left(1-\frac{L(L+1)}{2N}\right)\;.
\label{eq:pL}
\end{multline}
Note that eq.(\ref{eq:pL}) is valid for any $L\geq 1$ since $S_1\equiv 0$ and $c_2\equiv0$.

A similar expression can be written for the densities of loops of length $L$:
\begin{equation}
d_{L}^{loop}=\lim_{\mathcal{N}\rightarrow\infty}\frac{1}{\mathcal{N}}\frac{{\mathcal{N}}^{L}}{2L}\left[R_{L}(p_1,\dots,p_{L-1})+\frac{c_L}{{\mathcal{N}}^{L-1}}\right]\;,
\end{equation}
where again the function $R_L$ represents the probability of generating a loop of size $L$ by more than one BTS. Since the probability of having $k$ consecutive links is $O(\cN^{-k})$ the function $R_L$ is $O({\mathcal{N}}^{-L})$ and then
\begin{equation}
d_{L}^{loop}=\frac{c_L}{2L}=\frac{1}{N}\frac{z^{L}}{2L} \qquad \Longrightarrow \qquad
c_L=\frac{z^L}{N}\quad\text{for }L\geq 3\, .
\end{equation}
In other words, making a loop by randomly choosing smaller structures is more improbable than randomly generating directly such a loop.

The detailed computation of the coefficients $a_k$ from Eq.~(\ref{eq:pL}) is made in Appendix A. Here we just quote the result
\begin{align}
&a_1=z+\frac{1}{N}(2z^2-z)\, ,\\
&a_L=\frac{1}{N}(z^{L+1}-z^{L})\,\,\,\,\mathrm{for}\,\,L\geq 2\, .
\end{align}

Plugging these coefficients in Eq.~(\ref{eq:cavityF}) we finally get
\begin{equation}
f=\frac{z}{2}\left(1-\frac{1}{N}\right)\phi_1^a+\frac{z^2}{2N}\left(2\phi_1^a-\phi_2^a\right)+\frac{1}{N}\sum_{L=3}^{\infty}\frac{z^L}{2L}\left[\phi_L^c-L(\phi_L^a-\phi_{L-1}^a)\right] + \phi_{site}\, . \label{eq:F1cav}
\end{equation}
We observe that the sum on the r.h.s matches the sum over loops entering eq.\eqref{eq:F1}. Moreover, if the term $\phi_{site}$ is expressed by means of cavity fields, one finds exactly eq.\eqref{eq:F1}. This can be immediately seen in the case of zero external field in all the paramagnetic phase, where variables are unbiased and we have $\phi_{site}=-T(1-\ell)\log2$ with $\ell=z+(z^2/2-z)/N$ being the density of edges in the factor graph (i.e., the number of edges in the factor graph per variable node). Substituting this expression for $\phi_{site}$ in Eq.~\eqref{eq:F1cav}, simply gives:
\begin{equation}
f=-T\Big(\log2-\frac{z}{2}\mathbb{E}_J\log\mathrm{ch}(\beta J)\Big)+\frac{z}{2N}T\ \mathbb{E}_J\log\mathrm{ch}(\beta J)- \frac{T}{2N}\sum_{L=3}^{\infty}\frac{z^L}{L}\mathbb{E}_{\{J_i\}}\log\left[1+\prod_{i=1}^{L}\mathrm{th}(\beta J_i)\right]\, , 
\end{equation}
thus recovering the simplified replica result of Eq.~\eqref{eq:fh0}.

We can conclude that the replica calculation reproduces correctly all the topological structures involved in the $1/N$ corrections to the free energy density. Incidentally we note that self intersecting loops occur only with probability $N^{-2}$ and they do not contribute to $1/N$ corrections.

Let us finish this section by giving a different interpretation to the present results.
We have seen that under the assumption that finite size corrections can be computed by the cavity method in a graph with finite densities of certain topological structures, we have been able to reproduce the replica result (and give to it a more physical intuition).
However, we could assume that replica and cavity methods should provide the correct free-energy for a very large, but finite, system, and then conclude that the free-energy of a model only depends on the densities of certain topological structures.
This alternative view can be useful if one aims at computing the free-energy of a model defined on a finite dimensional lattice, by considering a lattice as a random graph with strong topological correlations, and making an expansion in these topological correlations (e.g., number of loops, but not only that).

\section{Numerical Analysis}
\begin{figure}
\includegraphics[width=\textwidth]{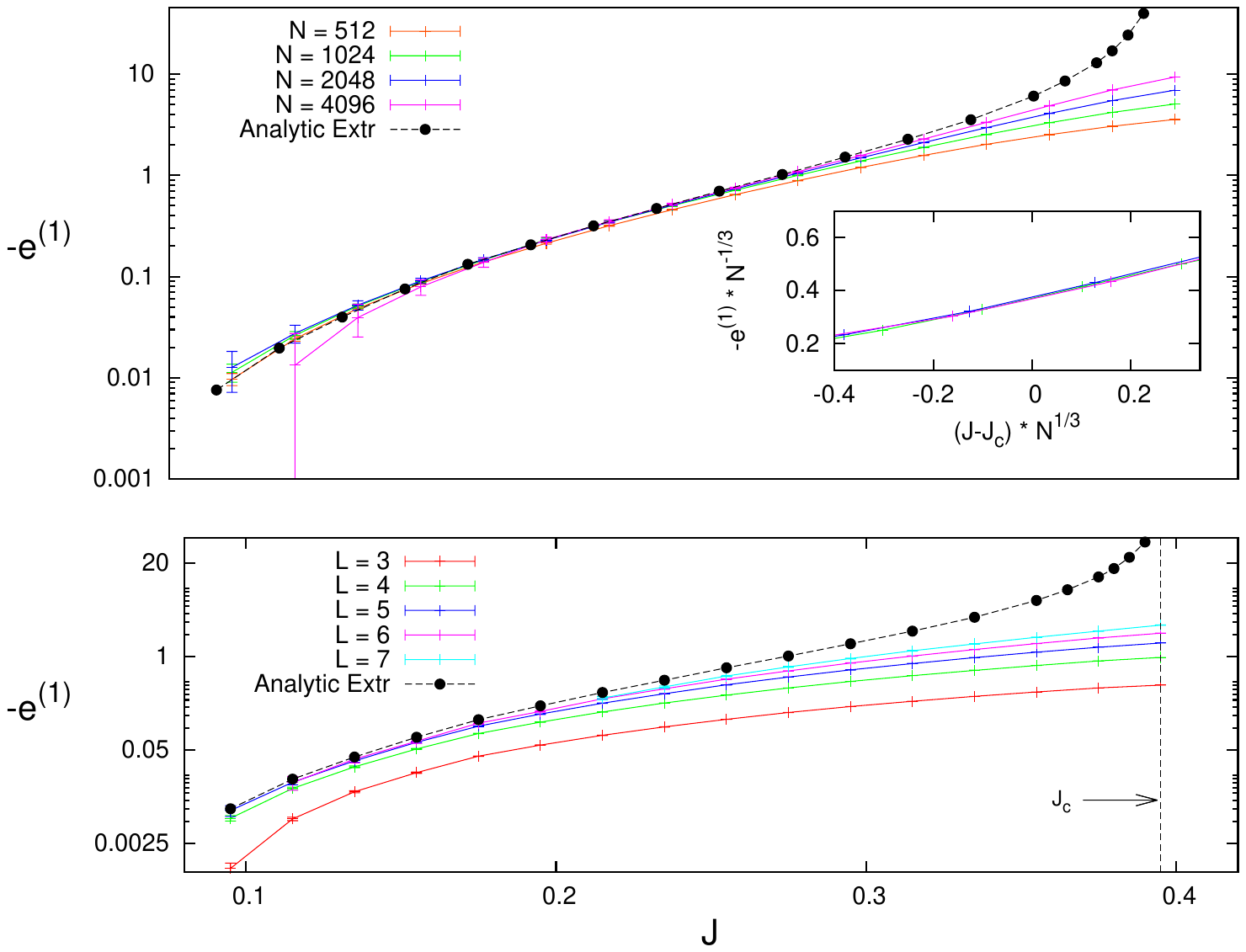}
\caption{Finite size corrections of the ground state energy density in the $T=0$ RFIM on Erd\"os-R\'enyi random graphs with mean connectivity $z=4$. In the upper panel we plot numerical data for different system sizes and the analytical formula given by eq. \eqref{eq:E1}. Close to the critical point (which is $J_c\approx0.395$) the scaling of the energy corrections is given by the mean field prediction \eqref{eq:E1meanfield} as confirmed by the data collapse shown in the inset.
In the lower panel we show the estimates of the formula \eqref{eq:E1}, truncating the sum over loops with a cutoff $L=3,4,5,6,7$ and extrapolating the whole series as explained in the main text. \label{fig:energycorr}}
\end{figure}

In this Section we check the validity of our analytical expressions for the free energy corrections, eq. \eqref{eq:F1}, against numerical simulations. Since from Monte Carlo simulations one obtains the energy of the systems, in order to avoid an integration in temperature we decided to work perform the simulations at zero temperature, where energy and free energy coincides.
Moreover since eq. \eqref{eq:F1} holds for arbitrary disorder in the interaction and in the external field, we choose to keep the former deterministic and the latter randomly distributed. In this case in fact an exact polynomial algorithm is available to calculate the ground state.
Therefore we apply eq. \eqref{eq:F1} to compute the finite size corrections to the energy density of the zero-temperature Random Field Ising Model (zt-RFIM) and compare with numerical simulations. The model is defined by the following Hamiltonian:
\begin{equation}
\mathcal{H}=-J\sum_{i,j}C_{ij}\sigma_{i}\sigma_{j}-\sum_{i}h_i\sigma_{i}\,\, ,
\end{equation}
where the random magnetic fields are Gaussian random variables of zero mean and variance $\overline{h_i^2}=1$ and the ferromagnetic exchange coupling $J$ take values in the interval $[0,\infty)$. The underlying graph topology is that of a  Erd\"{o}s-R\'enyi random graph. Due to the FKG\cite{FKG} inequality the model does not undergo replica-symmetry-breaking\cite{NoglassRFIM} at any value of the ferromagnetic interaction strength $J$, so that our formulae for the finite size (free) energy density corrections remain valid also below the critical point, provided that a single pure state is selected. In the ferromagnetic phase the existence of two energy minima generates additional finite size fluctuations, which are proportional to $N^{-1/2}$. These kind of \textit{interstate} fluctuations overcomes the $1/N$ \textit{intrastate} contribution, which becomes practically invisible in numerical experiments. In this work we compare analytical predictions and numerical results only in the paramagnetic phase $J<J_c$.\\
The uniqueness of the ground state of the model allows to translate the formula \eqref{eq:F1} for the free energy density corrections into the corresponding expression for the ground state energy density corrections. We write the ground state energy density as the leading term plus the $\ouN$ correction:
\begin{equation}
e^{\text{GS}}(N)=e^{\text{GS}}_0+\frac{1}{N}e^{(1)}+o\left(\frac{1}{N}\right)\ , \label{eq:eGS}
\end{equation}
 where $e^{(1)}$ reads:
\begin{equation}
e^{(1)}=-\left(z-\frac{z^2}{2}\right)\overline{|h^c|}-\frac{z}{2}e_1^a-\frac{z^2}{2}(e_2^a-2e_1^a)+\frac{1}{2}\sum_{L=3}^{\infty}\frac{z^L}{L}\left[e_L^c-L(e_L^a-e_{L-1}^a)\right]\,  . \label{eq:E1}
\end{equation}
The random variable $h^c$ is the cavity field, distributed according to the zero temperature solution of eq. \eqref{P(h)}, while $e^{a/c}_L$ are the energies of open and closed chains in the graph. The computational time cost of computing the energy density of a chain of size $L$ by enumeration is exponentially increasing in $N$, therefore only partial sums up to $L=7$ in eq. \eqref{eq:E1} have been considered in Figure \ref{fig:energycorr}. To accurately compute the whole $L$ series, especially near the critical point, some assumptions has to be made about the large $L$ behaviour of its term. Some of the authors have been developing a formalism through which a spectral representation  of the replicated transfer matrix \cite{Lucibello13,Martin1996, Janzen2010} can be obtained. Using this result the leading behaviour 
\begin{equation}
e_L^c-L(e_L^a-e_{L-1}^a)\sim A L \lambda^L
\end{equation} 
has been established for the zero temperature RFIM, which allows to analytically sum the remaining terms of the series (from $L=8$ to infinity). The coefficient $\lambda$ is given by the first eigenvalue of the replicated transfer matrix and gives the decay rate of ferromagnetic correlation functions. It can be computed to high precision with population dynamics techniques or as the first eigenvalue of an integral operator. The coefficient $A$ instead has been obtain from a fit of the first five point of the series. As an alternative approach assuming the validity of the $A L \lambda^L$ behaviour (which fares much better then a simple exponential decay assumptions) both $A$ and $\lambda$ could be inferred from a fit of the first terms of the sum. 
The finite size corrections of the energy in the RFIM at zero temperature diverges as $e^{(1)} \propto \frac{1}{ 1-z  \lambda}$, at odds with the double pole divergence $e^{(1)} \propto \frac{1}{(1-z  \lambda)^2}$ which can be found at finite temperature. This matter will be elucidated in a future work\cite{Lucibello13}.

\begin{figure}
\includegraphics[bb=150bp 50bp 702bp 550bp,scale=0.7]{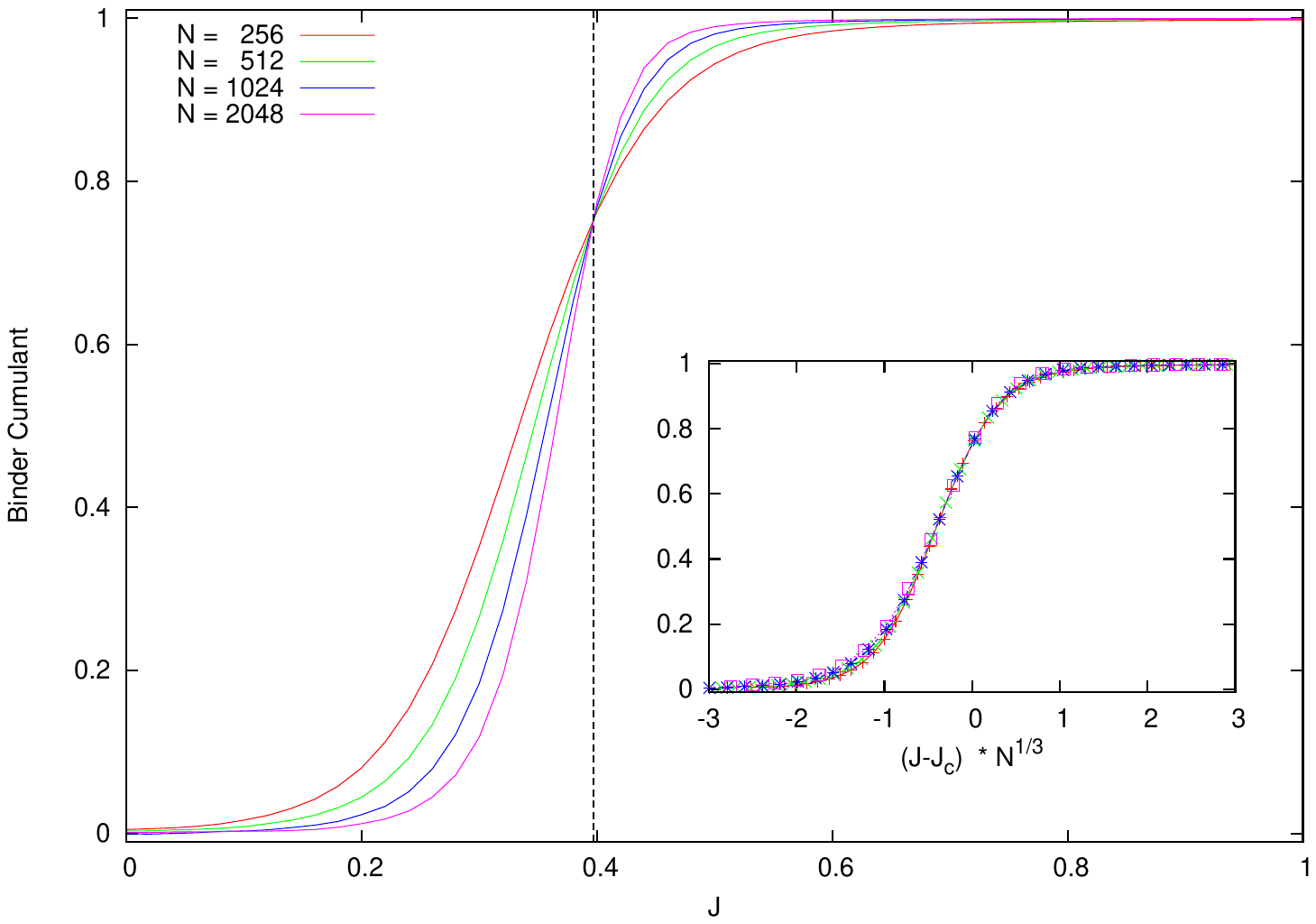}
\caption{Binder cumulant for the $T=0$ RFIM on Erd\"os-R\'enyi random graphs with mean connectivity $z=4$ for different system sizes as a function of the exchange interaction $J$. A vertical dashed line is drawn in correspondence of the critical point $J_c\sim 0.395$\,. In the inset it is shown the data collapse in the critical region using the scaling variable $(J-J_c)N^{1/3}$ for the reduced interaction. 
\label{fig:binder}}
\end{figure}

\begin{figure}
\includegraphics[width=\textwidth]{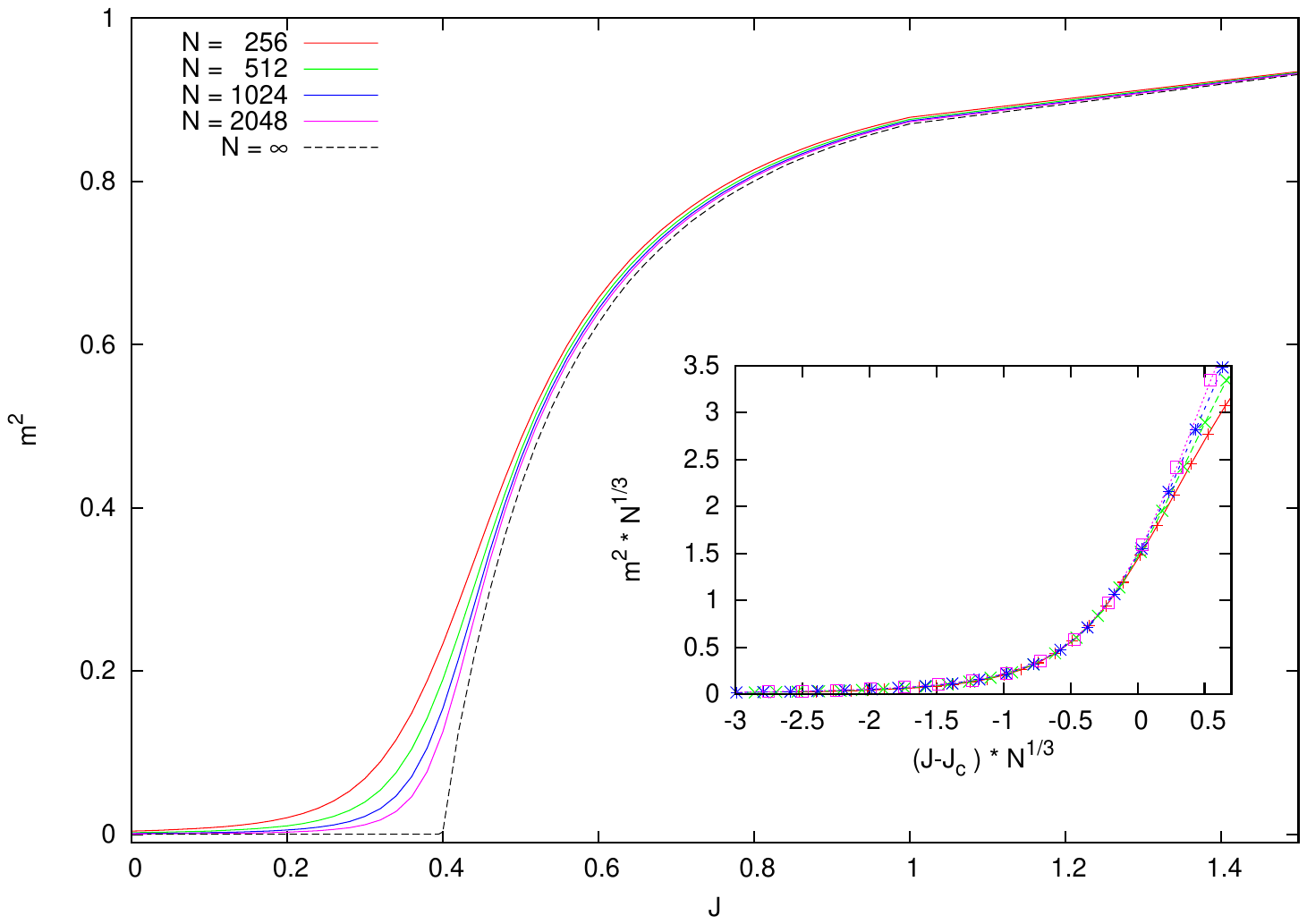}
\caption{Average squared magnetization for the $T=0$ RFIM on Erd\"os-R\'enyi random graphs with mean connectivity $z=4$ for different system sizes as a function of the exchange interaction $J$. 
In the inset it is shown the data collapse in the critical region using the scaling $N^{\frac{1}{3}}$ both for the reduced interaction and for the squared magnetization. 
\label{fig:mag2}}
\end{figure}

At the critical point a scaling analysis of the correction $e^{(1)}$ can be performed. Calling $\tau=|J-J_c|$ the distance from the critical point, mean field theory\cite{Natterman1998} predicts the following finite size scaling for $\tau$ and $e^{(1)}$ in the critical region:
\begin{align}
\tau &= \frac{\tilde{\tau}}{N^{1/3}}\, ,\\
e^{(1)} &=\tilde{e}^{(1)}N^{1/3}\, ,\label{eq:E1meanfield}
\end{align} 
The leading correction to the thermodynamic ground state energy density is of order $O(N^{-2/3})$ in the whole critical region:
\begin{equation}
e^{\text{GS}}(N)=e^{\text{GS}}_0+\frac{1}{N^{2/3}}\tilde{e}^{(1)}+o\left(\frac{1}{N^{2/3}}\right)\ \ \ \text{for}\ \ \ J\rightarrow J_c \ .\label{eq:eGScritic}
\end{equation}
Furthermore eq. \eqref{eq:eGS} is not valid in the ferromagnetic phase (for reasons mentioned in the beginning of this Section), where the leading correction happens to be of order $O(N^{-1/2})$:
\begin{equation}
e^{\text{GS}}(N)=e^{\text{GS}}_0+\frac{1}{N^{1/2}}e'^{(1)}+o\left(\frac{1}{N^{1/2}}\right) \ \ \ \text{for}\ \ \ J> J_c \ .\label{eq:eGSferro}
\end{equation}
The numerical experiment is performed on a Erd\"{o}s-R\'enyi random graph with average connectivity $z=4$.We compute
the ground state energy with the Minimum-cut algorithm \cite{HartNow99,MidFis02}, using the Lemon Library\cite{lemon}. To draw the profiles of the energy density corrections in Figure \ref{fig:energycorr} we took the average over $10^8$ samples for each system size. In the same figure we compare the numerical data with the analytical prediction given by eq. \eqref{eq:E1} and check the finite size scaling relation given by equation \eqref{eq:E1meanfield}. In Figure \ref{fig:binder} we report the Binder cumulant:
\begin{equation}
\text{Bi} = \frac{3}{2}\left[1- \frac{\overline{m^4} }{3\left(\overline{m^2}\right)^2}\right]\ ,
\end{equation}
 for system sizes ranging from $N=256$ to $N=2048$. From the intersection of the curves we identify the critical point, obtaining  $J_c\sim 0.395(1)$. \\
Figure \ref{fig:mag2} shows the behaviour of the averaged squared magnetization $\overline{m^2}$. The finite size scaling of $\overline{m^2}$ in the critical region is given by the following scaling relation:
\begin{equation}
\overline{m^2}=O(N^{-1/3})=O(\tau)\ \ \ \text{for}\ \ \tau\rightarrow 0.
\end{equation} 
This scaling form is confirmed by the data collapse shown in the inset of Figure \ref{fig:mag2}.\\

\newpage

\section{Summary and Conclusions}
In this work we performed a thorough analysis of the $\ouN$ correction to the free energy density in disordered Ising models defined on Erd\"{o}s-R\'enyi random graphs. We derived an analytical formula which can be easily used to quantify finite size effects, avoiding the subtleties associated with the diagonalization of the Hessian. We also checked the correctness of our results through a numerical study of the RFIM at zero temperature, and found that the finite size corrections to the ground state energy density are in perfect agreement with the analytical prediction.\\
More care has to be paid when studying finite size corrections in the ferromagnetic ordered phase. The formulae derived in this work are intended to be correct only when a single pure state is concerned, since they represent fluctuations inside a single pure state. Below the critical point, the continuous appearance of a couple of equivalent pure states, generates additional \textit{interstate} fluctuations, which cannot be described by the formula \eqref{eq:F1}. The nature of the low-temperature finite-size corrections is non-perturbative, so a different approach has to be taken in order to quantify them. Heuristic arguments and numerical simulations suggest that the first term of the free energy expansion is $O\left(N^{-1/2}\right)$ (at variance with the normal $\ouN$ behaviour), which dominates the \textit{intrastate} contribution.
Analogously when exponentially many pure states are involved, as in the case of spin glasses in their glassy phase, we expect the leading finite size correction to be much bigger than $\ouN$ the  and expressions \eqref{fRS} and \eqref{eq:F1} to not hold anymore. 
We also showed how replica results for the $1/N$ corrections to the free energy density can be derived also in the cavity formalism, resorting to an auxiliary graph ensemble which in some sense lifts the $\ouN contributions$ to the leading order. It would be interesting to see if this combinatorial derivation could be transposed to other graph ensembles.
Moreover we expect  our main result eq. \eqref{eq:F1} to hold some degree of universality, depending only on a few topological properties such as the mean residual degree $z$. 

\section*{Acknowledgments}
This research has received financial support from the European
Research Council (ERC) through grant agreement No. 247328, from the
Italian Research Minister through the FIRB project No. RBFR086NN1 and from the FP7 FET OPEN project
Enlightenment 284801.

\appendix

\section{Coefficients $a_L$}
The coefficients $a_L$ can be determined by computing recursively all the functions $S_L$.
A little bit of thought should convince oneself that $S_L$ is given by the probability of the event $E_L\equiv\{\text{L consecutive links are present}\}$, by splitting it in at least two smaller events $E_{L_1}$ and $E_{L_2}$ with $L_1+L_2=L$. Since $p_L=\mathrm{Prob}[E_L]$ it should be clear that $q_L\equiv p_L-S_L$ is like a ``connected'' probability to obtain the $L$ links from a unique structure. It is not hard to derive a recursive equation for the functions $S_L$, valid for any $L\geq 2$:
\begin{equation}
S_L=q_1 p_{L-1}+q_2 p_{L-2}+\dots+q_{L-1}p_1\, , 
\end{equation}
from which we get, for any $L\geq1$,
\begin{equation}
p_L(1+q_0)=\sum_{k=0}^{L}q_k p_{L-k}\, , 
\label{pl}
\end{equation}
where $q_0\equiv 0$ and $p_0=1$ thanks to the fact that $p_L=z^L[1-L(L+1)/2N]$. The above equation can be easily solved by introducing the generating functions:
\begin{equation}
\begin{aligned}
&p(x)\equiv\sum_{k=0}^{\infty}p_kx^k\ ,\\
&q(x)\equiv\sum_{k=0}^{\infty}q_kx^k\,\, , 
\end{aligned}
\end{equation}
that must satisfy
\begin{equation}
(1+q_0)[p(x)-p_0]=q(x)p(x)-q_0p_0\,\,\,\, \Longrightarrow \,\,\,\, q(x)=p_0-\frac{1}{p(x)}\, . 
\end{equation}
Keeping only terms up to order $1/N$ the result is 
\begin{equation}
q(x)=zx-\frac{1}{N}\frac{zx}{1-zx}\, , 
\end{equation}
implying
\begin{equation}
q_1=z\left(1-\frac{1}{N}\right)\,\,\,\,\,\,\, q_k=-\frac{z^k}{N}\,\,\,\,\mathrm{for}\,\,k\geq 2\,\,. 
\end{equation}
Rewriting eq.(\ref{pl}) as 
\begin{equation}
q_L=\sum_{k=L}^{\infty}(k-L+1)a_k+\sum_{k=L+1}^{\infty}c_k\,\,, 
\end{equation}
we can obtain
\begin{equation}
q_L-q_{L+1}=\sum_{k=L}^{\infty}a_k+c_{L+1}\,\,\,\, \Longrightarrow \,\,\,\, \sum_{k=L}^{\infty}a_k=q_L=-\frac{z^L}{N}\,\,\,\ \forall L\geq 2\,\, , 
\end{equation}
by noticing that $c_L=-q_L$ for $L\geq 3$. Moreover, for $L=1$ we have
\begin{equation}
\sum_{k=1}^{\infty}a_k=q_1-q_2=z+\frac{z^2-z}{N}\,\,. 
\end{equation}
In conclusion the coefficients $a_L$ are given by the following expressions
\begin{align}
&a_1=z+\frac{1}{N}(2z^2-z)\,\, ,  \\
&a_L=\frac{1}{N}(z^{L+1}-z^L)\,\,\,\, \mathrm{for}\,\, L\geq 2\,\, . 
\end{align}

\section{Combinatorics of $\tr \left(T^L\right)$}
\label{app:combin}
Here we prove eq. (\ref{eq:tracciacatene}), relating $\tr  \left(T^L\right)$ in the small $n$ limit to the free energies of open and closed cavity chains. Let's rewrite our $2^n\times 2^n$ matrix as:
\begin{equation}
T(\sigma,\sigma^{\prime})=z\,\mathbb{E}\left\{\frac{1}{[2\text{ch}(\beta h)]^n}\left[e^{J\sum_a\sigma_a \sigma_a' + h \sum_a\sigma_a'}-\frac{1}{[2\text{ch}(\beta h')]^n}\left(\int \dd\tau\, e^{J\sum_a\sigma_a \tau_a + h \sum_a\tau_a}\right)e^{h'\sum_a\sigma'_a}\right]  \right\}\, .
\end{equation}
where expectation is taken over the coupling $J$ and the cavity fields $h, h'$, which are distributed according to the solution of eq. \eqref{P(h)}. We immediatly note that the factor $[2\text{ch}(\beta h)]^n$ reduces to $1+n\log2\text{ch}(\beta h)+o(n)$, in the small $n$ limit, allowing to rewrite $T(\sigma,\sigma^{\prime})$, with $o(n)$ accuracy, as:
\begin{equation}
T(\sigma,\sigma^{\prime})=z\,\mathbb{E}\left[e^{J\sum_a\sigma_a \sigma_a' + h \sum_a\sigma_a'}-\left(\int \dd\tau\, e^{J\sum_a\sigma_a \tau_a + h \sum_a\tau_a}\right)e^{h'\sum_a\sigma'_a}\right] + n z \mathbb{E}_h\log2\text{ch}(\beta h)+o(n)\, .\label{eq:Tcomplete}
\end{equation}
We recognize that the term $\mathbb{E}[e^{J\sum_a\sigma_a \sigma_a' + h \sum_a\sigma_a'}]$ is the replicated transfer matrix of a one-dimensional chain, and so when we take the trace  of $T(\sigma,\sigma')$ we simply get:
\begin{equation}
\text{Tr}[T] = -n\beta z[\phi_1^c-\left(\phi_1^a+\beta^{-1}\mathbb{E}_h\log2\text{ch}(\beta h)\right)] + o(n)\,\, .
\end{equation}
When computing the trace of $T^L$, for $L\geq2$, the term $nz\mathbb{E}_h\log2\text{ch}(\beta h)$ in eq. \eqref{eq:Tcomplete} gives only contributions of order $o(n)$ and thus can be completely neglected in the following calculation. 
Let's call $A=\mathbb{E}[e^{J\sum_a\sigma_a \sigma_a' + h \sum_a\sigma_a'}]$ and $B=\mathbb{E}[\left(\int \dd\tau\, e^{J\sum_a\sigma_a \tau_a + h \sum_a\tau_a}\right)e^{h\sum_a\sigma'_a}]$. The product $T^L$ is formed of a linear combination of all the possible products of $L$ matrices chosen between $A$ and $B$, therefore we now consider the traces of such products. A simple inspection shows immediately the $\tr \left(A^L\right)$ is nothing else that the replicated partition function of a cavity loop, that is a closed chain of length $L$ embedded in a locally tree-like random graph. Consider instead a term with one insertion of the matrix $B$, $\tr A\ldots ABA\ldots$. Since $B$ is factorized, its insertion prevents the closure of the chain and we obtain the replicated partition function of an open cavity chain of length $L$. Generalizing the argument we can see that the trace of a product containing $k$ matrices $B$ yields the product of $k$ replicated partition functions of open chains, whose total lengths adds up to $L$. Since in the $n\downarrow 0$ limit products of partition functions become the sum of free energies, we can write
\begin{equation}
\frac{\partial}{\partial n}\tr \left(T^L\right) = -\beta z^L \big[\ \phi^c_L + \sum_{l=1}^{L} b_l\, \phi^a_l\ \big] + o(1)\, ,
\end{equation}
where the coefficients $b_l$ have to be determined. It is easy to see that $b_L = -L$ and $b_{L-1}=L$, while a simple combinatoric argument gives the remaining coefficients. We can construct an open chain of length $l<L-1$ in the first $l+1$ positions of the product and than multiply for the $L$ possible ways of obtaining the same trace. So we consider products of the form $BA^{l-1}B\times\{2^{L-l-1} \textit{ different combinations of A and B}\}$. Taking into account the number of insertions of $B$ in the last $L-l-1$ positions we obtain
\begin{equation}
b_l = L\times \sum_{k=0}^{L-l-1} (-1)^k {L-l-1 \choose k}= 0 \qquad \textit{for\quad } l < L-1\ ,
\end{equation}
which immediately yields eq. \eqref{eq:tracciacatene}.

\bibliographystyle{plain}
\bibliography{bibliography}

\end{document}